\documentclass[sn-vancouver]{sn-jnl}

\jyear{2022}%

\theoremstyle{thmstyleone}%
\theoremstyle{thmstyletwo}%

\theoremstyle{thmstylethree}%
\usepackage{ulem}
\usepackage{amsmath}

\newcommand{\oin}{\omega_{\rm{I}}}
\newcommand{\of}{\omega_{\rm{F}}}
\newcommand{\op}{\omega^\prime}
\newcommand{\omax}{\omega_{\rm max}}
\newcommand{\lvl}{\mathcal{L}_{\rm VL}}
\newcommand{\xp}{\bf{X}_{\mathcal{I}}}
\newcommand{\bom}{\boldsymbol{\omega}}

\begin{document}

\title[Geometry of Vortex Lines]{Geometric characterization of vortex lines in turbulence}


\author[1]{\fnm{Saumav} \sur{Kapoor}}\email{saumav.kapoor@icts.res.in}

\author[1]{\fnm{Rama} \sur{Govindarajan}}\email{rama@icts.res.in}

\author*[1]{\fnm{Siddhartha} \sur{Mukherjee}}\email{siddhartha.m@icts.res.in}

\affil[1]{\orgname{International Center for Theoretical Sciences, Tata Institute of Fundamental Research (ICTS-TIFR)}, \orgaddress{\street{Survey No. 151, Shivakote}, \city{Bengaluru}, \postcode{560089}, \state{Karnataka}, \country{India}}}

\abstract{Vorticity in turbulent flows is often organized into complex geometries that influence the dynamics. We use a relatively novel approach to describe these geometries: that of obtaining segments of vortex lines embedded in the flow. This enables us to quantify the geometric features of these objects. Vortex lines differ widely in their behaviour but we find some unifying features. The decay from high levels of vorticity is shown to happen over a short fraction of the vortex length. The local curvature is inversely related with the vorticity magnitude. Strong parts of vortex lines bundle together. It is hoped that this first work will generate interest in such quantification and the physics of vortex dynamics in turbulence.}




\maketitle

\section{Introduction}\label{intro}
Vortical coherent structures have asserted themselves as the ``sinews''~\cite{moffatt1994stretched} of Navier-Stokes turbulence~\cite{Saffman1973,she1990intermittent,jimenez1998characteristics,hussain1986coherent,hunt1988eddies}. 
Roughly, they may be described as intensely swirling regions of fluid surrounding cores with large vorticity magnitude. These structures have typical lengths of the order of the Taylor microscale while their cross-sections could be of the Kolmogorov lengthscales or smaller~\cite{jimenez1998characteristics}. Interestingly, they may be much longer-lived than the Kolmogorov timescale. While these structures are closely tied to small-scale intermittency~\cite{frisch1995turbulence,laval2001nonlocality}, they are intriguingly also responsive to the large-scales~\cite{vela2021synchronisation}. Their presence makes the modeling or approximate closure of turbulence theories a major challenge~\citep{paladin1987anomalous,frisch1995turbulence,dubrulle2019beyond,dubrulle2022multi}. They often represent small-scale extreme events both in Lagrangian~\citep{scatamacchia2012extreme,bitane2013geometry,saw2014extreme} and Eulerian~\citep{schumacher2010extreme,buaria2022vorticity} measurements due to the very large spatial gradients they support. They are also being identified as mediators of singular dissipation~\citep{debue2021three,faller2021nature}.

The existence of intense coherent vortices may be traced to the fact that vortex-stretching is more common than vortex-compression in most single-fluid turbulent flows, which is synonymous with a net positive enstrophy production~\cite{tsinober2009informal} (while in multiphase flows, with elastic interfaces interacting with turbulence, the situation may be different~\cite{mukherjee2019droplet}). Most of the high amplitude coherent vortices tend to organize around \textit{jet-like} vortex lines~\cite{she1990intermittent,she1991structure,sid2022} and the velocity structure is a superposition of background and self-inducing fields~\cite{sid2022}. Moreover, the structure of the vortex lines can be curved, twisted, under torsion, tangled and linked~\cite{moffatt1969degree,saffman1995vortex,xiong2019identifying}. These topologies are not exotic, and are in fact the norm in a turbulence field. An important aspect of these curved topologies is that their neighborhoods are often regions of high helicity density, total helicity being conserved for ideal fluids, alongside energy~\cite{moffatt1992helicity,moffatt2014helicity}. Helicity, which is the volume integral of $\boldsymbol{u}\cdot\boldsymbol{\omega}$ (the alignment of the velocity $\boldsymbol{u}$ with the vorticity $\boldsymbol{\omega}$ is also known as flow Beltramization), is directly linked to the suppression of non-linearity in the Navier-Stokes equation $(\boldsymbol{u}\times \boldsymbol{\omega})$~\cite{kraichnan1973helical,moffatt2014helicity}, which hinders the inter-scale energy transfer and acts as a self-attenuating mechanism~\cite{buaria2020self}. Under inviscid evolution (by the Euler equations), the knot structure (topology) of vortex lines is preserved, which is a direct consequence of Helmholtz laws of vortex motion, which both forbid vortex lines from ever crossing and preserve the flux of vorticity, which also implies helicity conservation~\cite{moffatt1969degree}. At finite Reynolds numbers, however, viscosity-mediated vortex reconnections~\cite{kida1994vortex,kerr1989simulation} take place, which form a pathway for energy dissipation and depletion of helicity~\cite{kimura2014reconnection,yao2020physical,zhao2021helicity,yao2022vortex}. 

We propose that all of this points to the need for identifying vortical structures as spatially-finite objects with extended geometrical and topological features that are key for illuminating some aspects of turbulence dynamics. This requires using tools that make these geometrical aspects tractable and fully quantifiable. For instance, common tools that use point criteria based on the velocity-gradient tensor and its invariants~\cite{hunt1988eddies,chong1990,jeong1995identification,dubief2000coherent}, appear limited in this context. More recently, headway has been made into describing tangled vortex tubes using vortex surface fields~\cite{xiong2019identifying}, as well as studying the distributions of vorticity extrema measured along space-filling vortex lines~\cite{Wang2012}. We develop a similarly motivated strategy for identifying vorticity (or ``vortex'') lines, henceforth referred to as VLs, which are defined at any given time by the trajectories of the dynamical system
\begin{equation}
    \frac{d\boldsymbol x(s)}{ds}=\boldsymbol{\omega}(\boldsymbol x(s), t),
\end{equation}
where $\boldsymbol{\omega}(\boldsymbol{x}, t)$ is the vorticity field at the spatial location given by $\boldsymbol x$ at time $t$. By definition, this makes a VL a curve, lying along $s$, which is everywhere tangent to the vorticity vector. It can, especially when closing on itself, be referred to as a ``vortex tube". It is important to distinguish this object from a ``line vortex", which is a vortex with a well defined notion of a cross-section as well as a circulation $\Gamma$. In inviscid flow, the identity of a line vortex can be retained in time with circulation remaining constant, so that the two notions could be identical. But in viscous flow, and particularly in turbulence, there is no clear definition of vortex cross-sections, especially when the cross-section of a vortical patch becomes large. Due to viscous diffusion, there is no guarantee of $\Gamma$ being a constant in time. Moreover, when the cross-section of a vortical patch becomes large, it is not possible to distinguish vorticity contributions from different ``line vortices", leading to their identities becoming blurred. A VL, on the other hand, is a well-defined object. In this work (as detailed in the Methods section), we shall identify VLs as curvilinear segments with their two ends being at prespecified high and low levels of vorticity magnitude. The latter serves as a cutoff since VLs are difficult to distinguish in regions where the vorticity magnitude drops to near zero. In contrast, Wang~\cite{Wang2012} instead track VL segments using \textit{extrema} of vorticity measured along VLs, and only study the path length and vorticity difference between subsequent extrema. Going a step further, our method retains information on vorticity along VLs, allowing greater control on subsequent analysis of different geometrical aspects along VLs vis-a-vis the local vorticity.

The object of this work is to develop a framework for studying the rich geometrical aspects of vortical coherent structures. Given our construction of the VLs, that start at relatively high levels of vorticity and meander through space until decaying to mild vorticity levels, regions with intense vorticity will invariably form small segments along a subset of VLs. Tracking these spatial curves will give us understanding on their geometrical properties, e.g., their total length, variations in curvature along their length, torsion, as well as physical quantities like velocity, vorticity and helicity along each VL. We also analyze persistence along VL segments, and show how vortical structures are organized. While we work with data from high resolution direct numerical simulations of homogeneous, isotropic turbulence, these techniques will readily extend to experimental data as well as to other kinds of flows including shear and wall-bounded flows.

\section{Data and Methods}\label{method}
\begin{figure}[!ht]
\centering
\includegraphics[width=\linewidth]{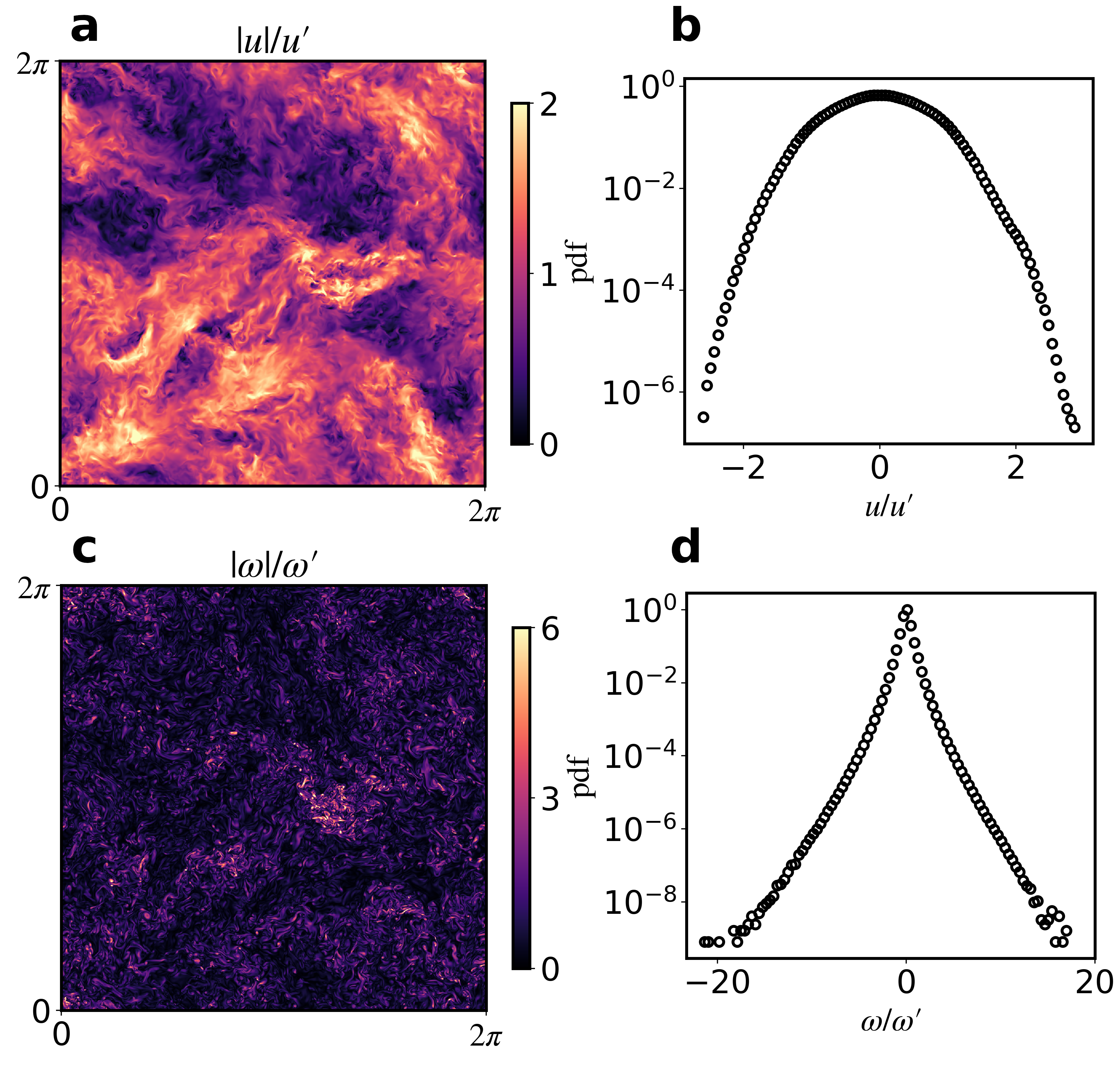}
\caption{Representative snapshots and distributions from the Johns Hopkins Turbulence Dataset that is used in this study. Panels (a) and (c) show a crossection of the velocity and vorticity magnitude fields, normalized by $u' = \langle \boldsymbol{u}\cdot\boldsymbol{u} \rangle^{1/2}$ and $\omega' = \langle \boldsymbol{\omega}\cdot\boldsymbol{\omega} \rangle^{1/2}$, respectively. The colorbars have been capped for better visualization. Panels (b) and (d) show the distribution of the velocity and vorticity, taken over all three directions.}
\label{fig:JHTD}
\end{figure}

We use a standard turbulence field snapshot, at an arbitrary time, from the Johns Hopkins Turbulence Dataset of homogeneous isotropic turbulence~\cite{perlman2007data,li2008public}. The simulation was performed on a $(2\pi)^3$ domain discretized on a $1024^3$ grid, with a viscosity $\nu = 0.000185$, total kinetic energy $E_{\rm tot} = \langle \boldsymbol{u}\cdot\boldsymbol{u} \rangle/2 = 0.705$ and rate of energy dissipation $\epsilon = 0.103$. This leads to the Taylor microscale $\lambda = \sqrt{15\nu u'^2/\epsilon} = 0.118$, Taylor-scale Reynolds number ${\rm Re}_{\lambda}=418$ and Kolmogorov lengthscale $\eta = (\nu^3/\epsilon)^{1/4} = 0.0028$ (all parameters are non-dimensional). We obtain the vorticity field from the velocity data using second order central differencing. We present representative snapshots of the kinetic energy and vorticity fields from this data, together with the velocity and vorticity distributions for reference, in Fig.~\ref{fig:JHTD}.

\subsection*{Vortex Line Identification}
To identify VLs, we simply track the path along a VL starting from a chosen high vorticity magnitude $\oin$ and marching, at each point following the direction of the local vorticity vector, to a final magnitude $\of$. We then start at the same starting point and march in the other direction, i.e., against the direction of the vorticity vector, till we reach $\omega \equiv
\sqrt{\boldsymbol{\omega}\cdot\boldsymbol{\omega}} = \of$. This method allows us to find all VLs connecting different spatial regions containing the predetermined levels of vorticity magnitude. 

We start by finding the set of all  locations $\xp$ in space where the vorticity magnitude $\omega $ is within a small fraction $f$ of a chosen initial threshold $\oin$, such that for all $\bf{x} \in \xp$ we have $\oin(1-f) < \omega({\bf{x}}) < \oin(1+f)$. We randomly choose $N$ points from the set $\xp$ to get ${\bf{X}}_\mathcal{O} \subset \xp$, and we use this subset of points ${\bf{x}}_0 \in {\bf{X}}_\mathcal{O}$ as the initial locations for tracking VLs. The forward trajectory of a VL starting from any ${\bf{x}_0} $ is obtained as a sequence of points $\{{\bf{x}}_k\}$ upon iterating ${\bf{x}}_{k+1} = {\bf{x}}_k + \bom({\bf{x}}_k) \Delta s$, where $\Delta s$ is the integration step. A similar operation is performed for the backward trajectory, at each point marching along $-\bom$. We stop the integration once the vorticity magnitude $\omega({\bf{x}}_{k+1}) \leq \of$. For ease of defining different vorticity levels, we normalize the vorticity field by the root-mean-square vorticity $\op = \langle \boldsymbol{\omega}\cdot \boldsymbol{\omega}\rangle ^{1/2}$ where $\langle \rangle$ denotes spatial averaging. We fix $\Delta s = 0.05$, which is 20 times smaller than the size of a single grid, which ensures that the vorticity-weighted step remains within a grid size even for the spatial points with extreme vorticity. Vorticity vectors at off-grid locations are calculated via trilinear interpolation. 

We devised this method to track VLs connecting specified vorticity thresholds, and in particular, to allow us to study how vorticity decays along VLs, how this variation reflects in the geometry of the VLs as measured by their curvature, and to enable the collection of statistics over a family of VLs that connect comparable regions of the flow field. 

\section{Results}\label{results}
\begin{figure}[!ht]
\centering
\includegraphics[width=\linewidth]{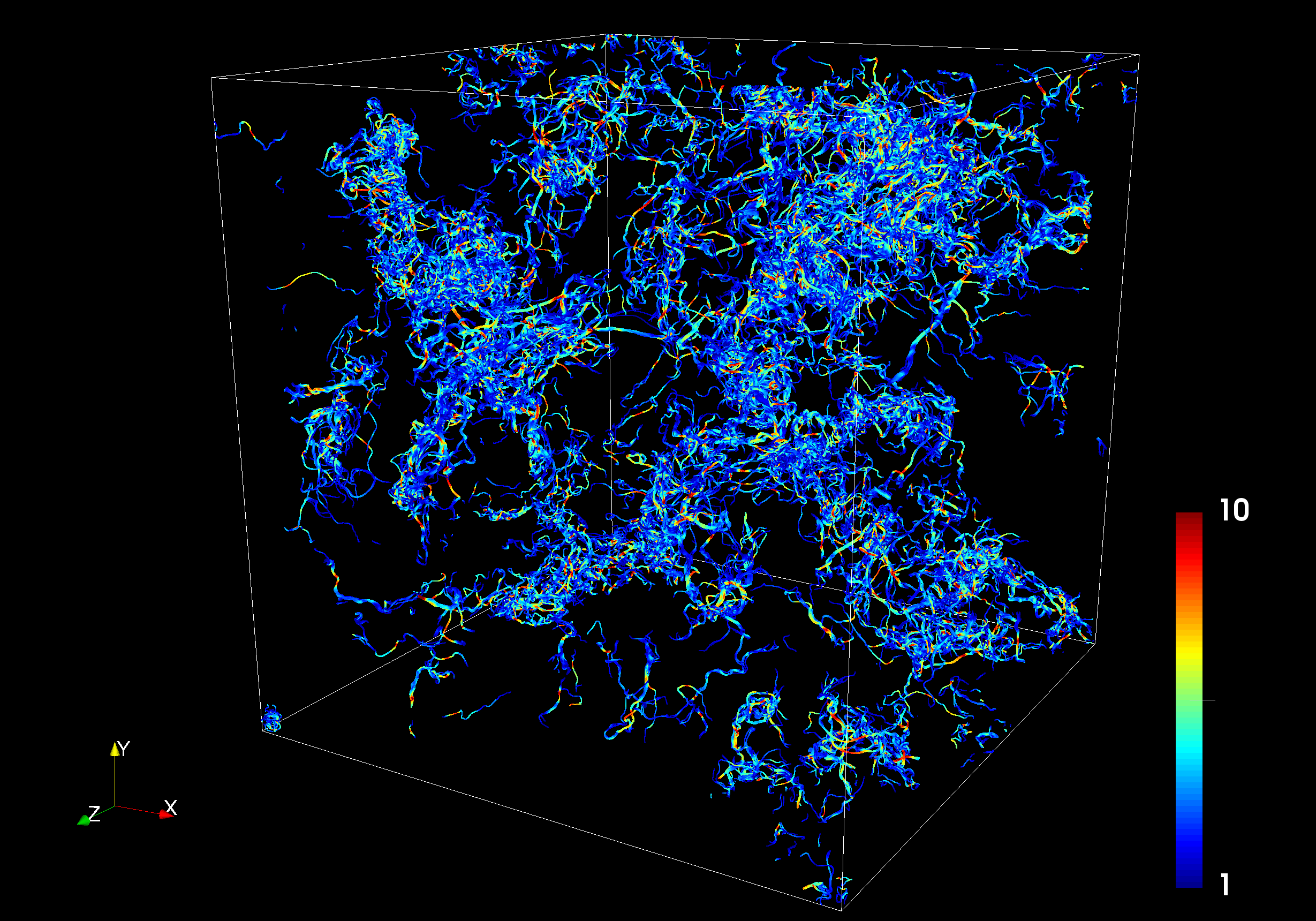}
\caption{A randomly chosen subset of VLs going from $10\op$ to $\op$, coloured by the vorticity magnitude. A highly anisotropic spatial distribution, typical of intermittency, is seen. The VLs seem to cluster densely and have very complicated geometrical features.}
\label{fig:VLsVolume}
\end{figure}

We begin with a compelling visual impression of the complex organization of VLs in the entire turbulence field. Fig.~\ref{fig:VLsVolume} shows a subset of $10000$ randomly chosen VLs originating from $\oin = 10\op \pm 0.1\op$ and ending at $\of=\op$, coloured by the vorticity magnitude along the lines. The intermittent nature of strong vorticity is immediately apparent, as the VLs fill space in a highly anisotropic manner with large pockets of quiescent regions interspersed between small and densely packed vortical regions (by contrast, a random sampling of a uniformly distributed quantity would be uniform). Identifying these VLs seprately allows us to characterize their variation in space, together with the variation of physical quantities like vorticity magnitude or curvature along these lines.

\begin{figure}[!ht]
\centering
\includegraphics[width=0.65\linewidth]{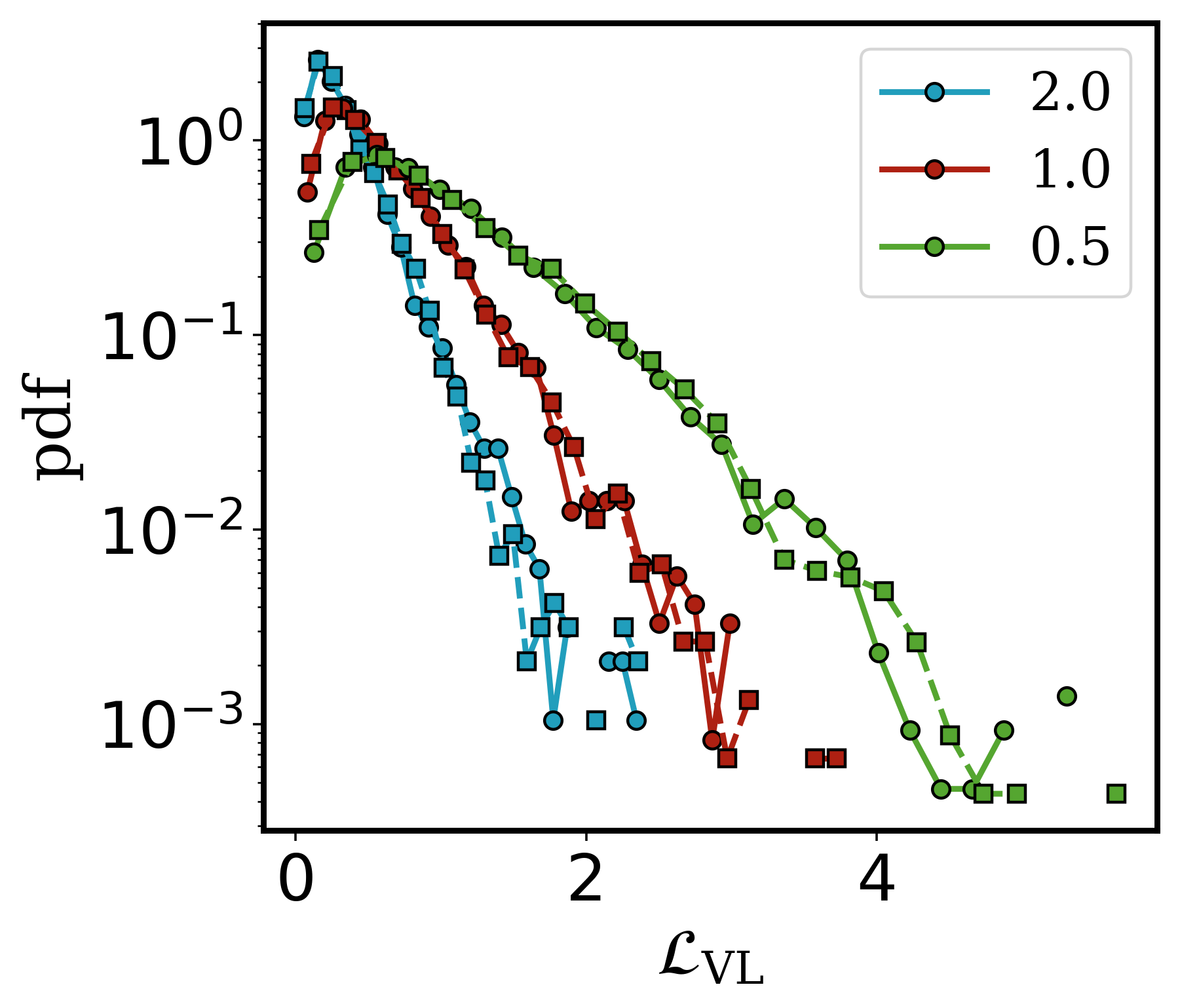}
\caption{Probability distribution of VL lengths, $\lvl$, for $\oin = 10\op$ (with tolerance fraction $f=0.1$), for different values of the final vorticity $\of = 2\op, \op$ and $0.5\op$. Circles (squares) show forward (backward) paths along VLs starting from $\oin$. Here $\lvl$ is presented relative to the edge of the simulation domain being $2\pi$.}
\label{fig:PDFVortLen}
\end{figure}

We first consider the distribution of VL lengths, $\lvl$, which we define as the total length from $\oin$ to $\of$, along each VL, in the forward or backward direction. Treating the directions separately allows us to check for any inherent fore-aft asymmetry that may be present in the VLs. Fig.~\ref{fig:PDFVortLen} shows, on a semi-log plot, the distribution of $\lvl$ for $\oin = 10\op$ (while fixing the initial tolerance fraction $f=0.1$) and several values of $\of \in \left \lbrace 2\op, \op, 0.5\op \right\rbrace$, calculated from $10000$ VLs. We find that $\lvl$ decays exponentially beyond an initial peak, while the slope of the decay becomes shallower for higher values of $\of$. Such a change in slope is to be expected, since all VLs extending up to a small $\of$ will contain as their subsets VLs extending up to larger $\of$, and will obviously be longer than their subsets. This becomes clear upon considering equally likely path-lengths $\lvl$ for $\of = 0.5\op$ in comparison to $\of = 2.0\op$, the former are almost double in length. This shows us that drop in vorticity is more rapid when the starting vorticity is higher. We shall quantify this observation further subsequently, especially in the context of non-monotonic changes in vorticity along VLs. We further show that there is no asymmetry between the forward and backward path lengths.

A similar observation of exponentially distributed VL lengths, obtained using path lengths between subsequent vorticity extrema, was obtained by Wang~\cite{Wang2012}. We have extended their result by showing that the distribution is exponential as long as $\oin\gg\of$. The distribution of $\lvl$ to the right of its peak remains exponential for different choices of $\oin$ (not shown) and $\of$. However, we note that the slope of the line in the PDF is much more sensitive to variations in $\of$ than variations in $\oin$, supporting the observation that the rate of drop of vorticity along VLs is much faster for higher vorticity.

As we laid down at the outset, the reason we wish to identify vortex lines individually is to study aspects of their coherence and persistence spatially (and later temporally). Moreover, we are interested in finding regions with clear signs of geometrical and structural complexities like twist, torsion, writhe and perhaps even links. These regions of the vorticity field are prone to vortex reconnections, and are thought to undergo a cascade of successively simplifying topological changes~\citep{kleckner2013creation,kleckner2016superfluid}. This process, moreover, generates multiple scales of motion and has been further linked to the ``turbulence cascade''~\cite{yao2020physical,yao2022vortex}.

There is a qualitative understanding that regions of high vorticity magnitude form \textit{jet-like} structures, which typically are straighter segments of the VLs, but to our knowledge this has not been quantified before. We perform such a quantification by obtaining a joint distribution between the curvature $\kappa$  and the vorticity magnitude.  We measure $\kappa$ along the VLs as
\begin{equation}
 \kappa = \frac{\sqrt{ (x_s y_{ss} - y_s x_{ss})^2 + (y_s z_{ss} - z_s y_{ss})^2 + (z_s x_{ss} - x_s z_{ss})^2 }} {{(x_s}^2 + {y_s}^2 + {z_s}^2)^{3/2}}
\end{equation}
where the subscripts refer to differentiation along the VL coordinate $s$.
\begin{figure}[!ht]
\centering
\includegraphics[width=\linewidth]{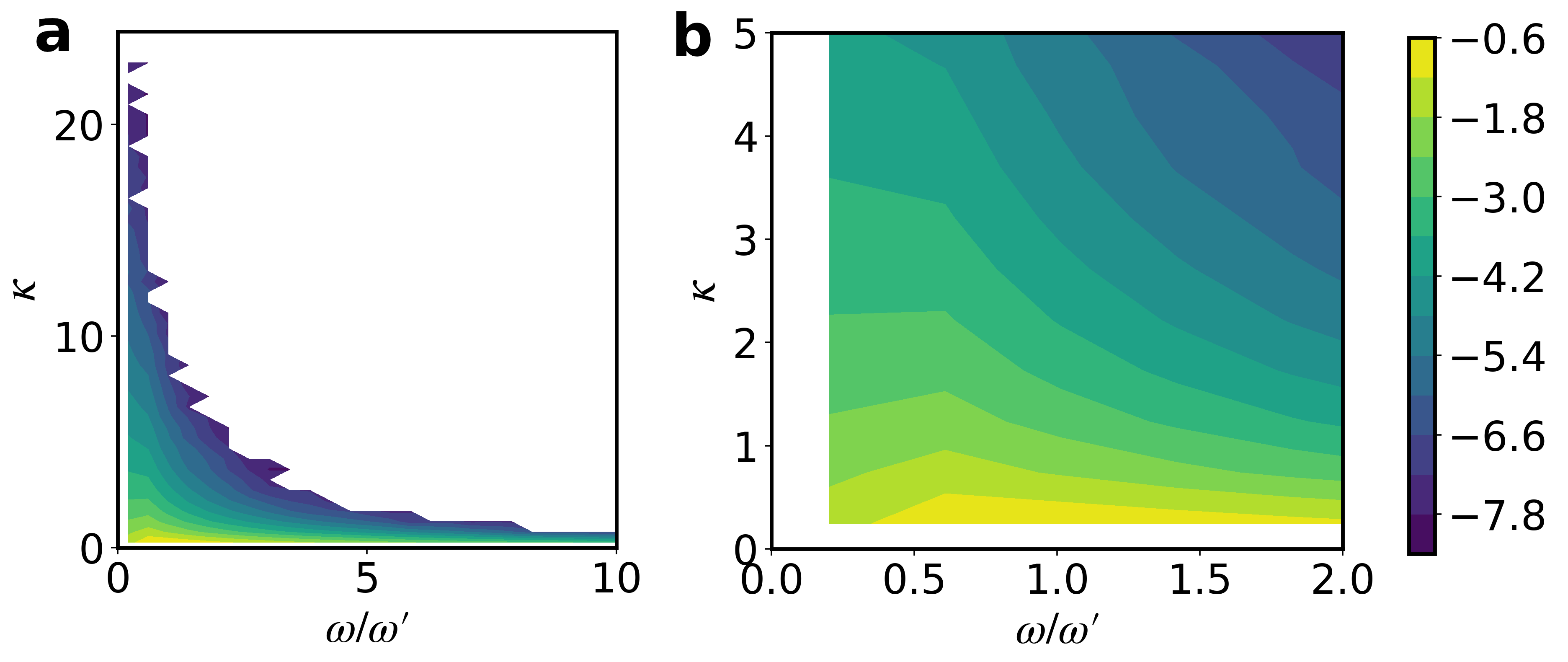}
\caption{(a) Joint-distribution of the curvature $\kappa$ and vorticity magnitude $\omega$ along VLs. Regions of high vorticity consistently have lower curvatures, while regions of mild to low vorticity can have a wide range of curvature from $0 <\kappa <20$. (b) A closer look at the peak of the distribution occurring at low $\omega$ and $\kappa$ values. The colorbar represents probability levels that are logarithmically spaced.}
\label{fig:JPDFCurvVort}
\end{figure}
Fig.~\ref{fig:JPDFCurvVort} shows this joint-distribution between $\kappa$ and $\omega$. We find a strong anti-correlation between high vorticity magnitudes and high curvature. This reflects that regions of high $\omega$ along VLs invariably have lower curvatures, while regions with mild to low $\omega$ can be anything from straight to highly convoluted, i.e., assume a wide range of $\kappa$ values. The former is likely to be a consequence of the vortex-stretching mechanism responsible for increasing $\omega$. Stretching occurs as a consequence of strain self-amplification and enstrophy production from an alignment between the straining eigen-direction and the local vorticity vector~\cite{tsinober2009informal}. This process stretches flux tubes while concentrating them together in their orthogonal plane, which would then increase the vorticity magnitude in the core region of VL-bundles. Invariably, this leads to the VLs becoming straight with typically low curvature. In fact, the highest vorticity magnitudes align into bundles displaying low curvature as we shall see below. To confirm this hypothesis, we need to track the dynamics of VLs in time together with strain-fields, a study we intend to carry out in the future.

\begin{figure}[!ht]
\centering
\includegraphics[width=\linewidth]{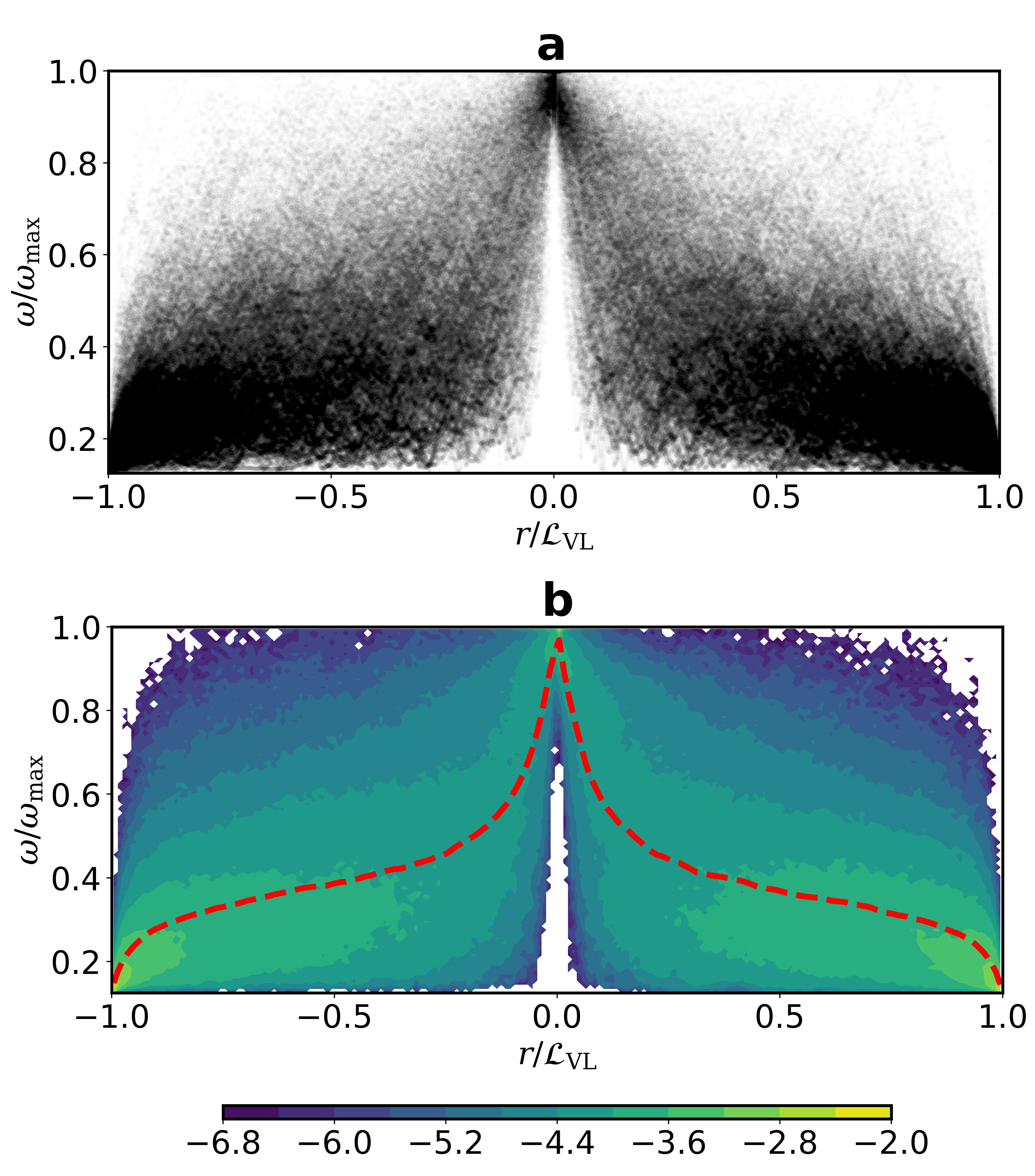}
\caption{(a) Vorticity magnitude is plotted along the length of VL segments about the vorticity local maxima, $\omax$ centered at $r=0$. The distance $r$ is scaled by $\lvl$ (measured separately for the forward and backward directions), such that at $r/\lvl = \pm 1$, $\omega/\omax =\beta= 1/8$. (b) Corresponding joint-distribution of $\omega/\omax$ and $r/\lvl$, showing the wide variation in the behaviour of VL segments. The red-dashed line shows the mean $\omega$ at each $r$, and the colorbar indicates logarithmically spaced probability levels.}
\label{fig:Decay}
\end{figure}

\begin{figure}[!ht]
\centering
\includegraphics[width=\linewidth]{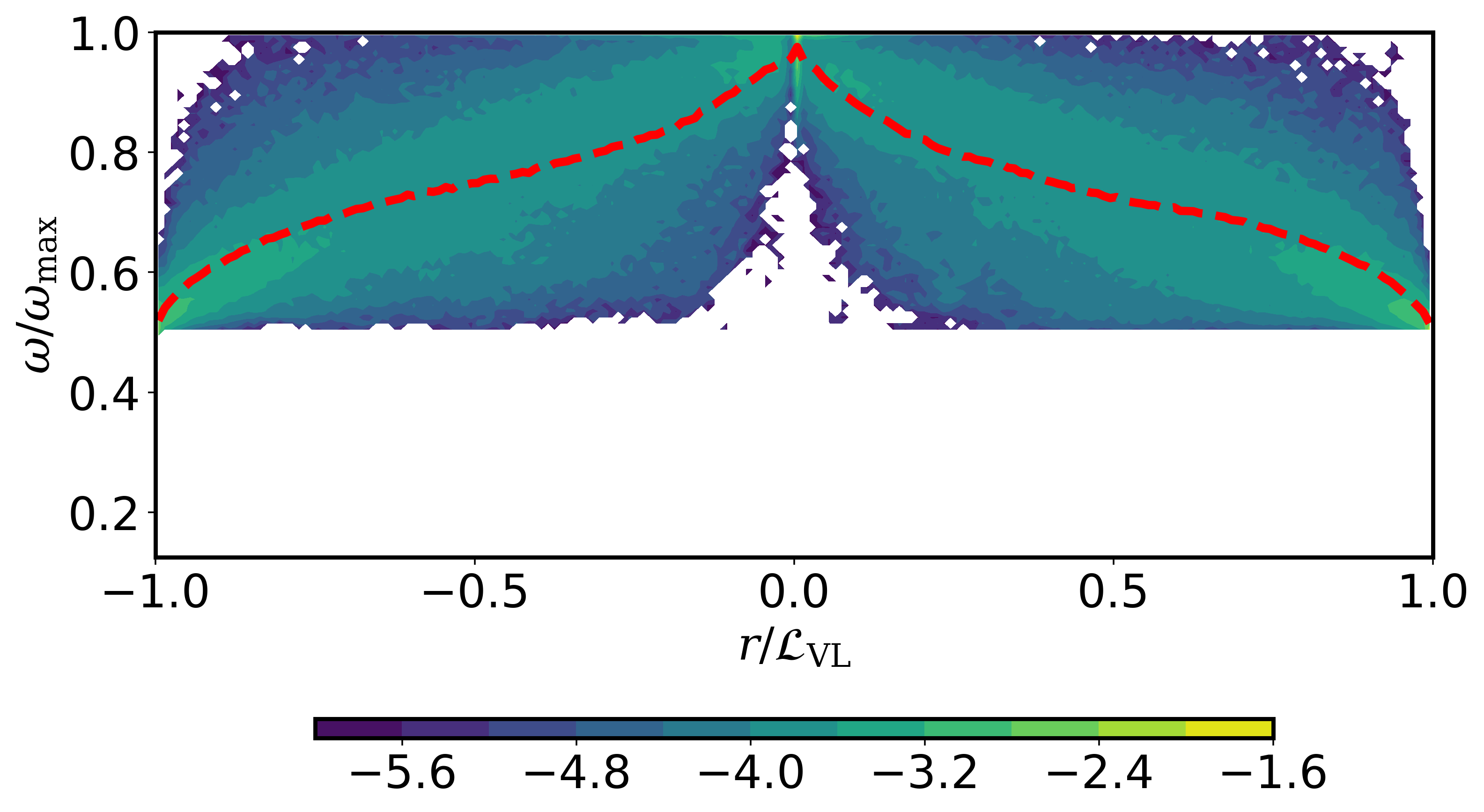}
\caption{Joint-distribution of $\omega/\omax$ and $r/\lvl$, with $\lvl$, for the first vorticity halving with $\beta=1/2$. The red-dashed line shows the mean profile, which confirms that the first vorticity halving is highly monotonic, and that the final approach to $r/\lvl \to \pm1$ is again rapid, due to the construction of the VLs.}
\label{fig:Decay2}
\end{figure}

Since each VL is a single coherent unit, the persistence of vorticity with a high amplitude along these lines is easy to obtain and forms an interesting metric useful for classification and checking for universality. In other words, we may ask whether all regions of high vorticity magnitude have similar structure. For this, we again consider a random sampling of 10000 VLs connecting $\oin = 10\op \pm 0.1\op$ to $\of = \op$. On every VL, we then find the point where the vorticity magnitude is the highest, at $\omega_{\rm max}$, which is often larger than $10\op$. From the point of maximum vorticity, we march along the VL (staying on coordinate $s$), both in the forward and backward directions, following the decrease in $\omega$ with distance $r$ until it reaches a specified fraction of $\omega_{\rm max}$. As seen earlier in Fig.~\ref{fig:PDFVortLen}, these lengths are exponentially distributed, so we normalize the distance $r$ with $\lvl$ (which is measured separately for the forward and backward directions), defined as the distance in either direction until $\omega/\omega_{\rm max} \leq \beta$ where $\beta$ is a specified fraction. In other words, at $r=0$ we have $\omega/\omax = 1$, and as $r/\lvl \to \pm 1$, $\omega/\omax \to \beta$. 

Fig.~\ref{fig:Decay}(a) shows the vorticity persistence profiles for a sampling of 1000 VLs, with $\beta = 1/8$. The measurement in the forward (backward) direction is shown along a positive (negative) $r$-axis. The level of opacity reflects the concentration of lines. Clearly, the ensemble of VLs have a symmetric persistence profile about $\omax$, and many VLs go through a sequence of local maxima decorating their overall decreasing trend. Fig.~\ref{fig:Decay}(b) show a joint-distribution between $\omega/\omax$ and $r/\lvl$, from all 10000 VLs, which shows that while there is a concentration of VLs decaying with a profile close to the mean (shown by the dashed red line), there is also a lot of deviation from the mean profile as well as non-monotonic variation of vorticity within the ensemble. The decay rate of the mean profile is expected to depend mainly on two factors, namely, the decay rate of vorticity along VLs and the amount of non-monotonicity and fluctuation of vorticity along VLs. High levels of non-monotonicity and fluctuation are expected to flatten the mean profile. Observing the mean profile in Fig.~\ref{fig:Decay}(b) shows, on an average, the vorticity at the level of $\omax$ persists for a very small distance and falls rapidly to half its value by $r/\lvl\approx 0.1$, reaffirming the previous result about high levels of vorticity decaying more rapidly. The subsequent halving from $\beta = 1/2$ to $\beta = 1/4$, however, proceeds much more slowly, which means that mean vorticity persists over longer lengths in this range of values, due to the fluctuating and non-monotonic variation of $\omega$. Interestingly, the decay from $1/4 \leq \omega/\omax \leq 1/8$ is again rapid. However, this is purely due to monotonic decrease of vorticity near $r=\lvl$, guaranteed by the construction of these VL segments. This observation is buttressed by the mean vorticity profile in Fig.~\ref{fig:Decay2}, which is restricted to the first vorticity halving. We find that there is little flattening, especially in comparison to the mean profile in Fig.~\ref{fig:Decay}(b). This affirms that the drop in vorticity during the first halving is not only rapid, but also highly monotonic. While it might be tempting to find the function that fits the mean profile, we refrain from doing so, since the variation within the ensembles is rich and it would be misleading to classify the VLs as having a universal profile.

\begin{figure}[!ht]
\centering
\includegraphics[width=0.9\linewidth]{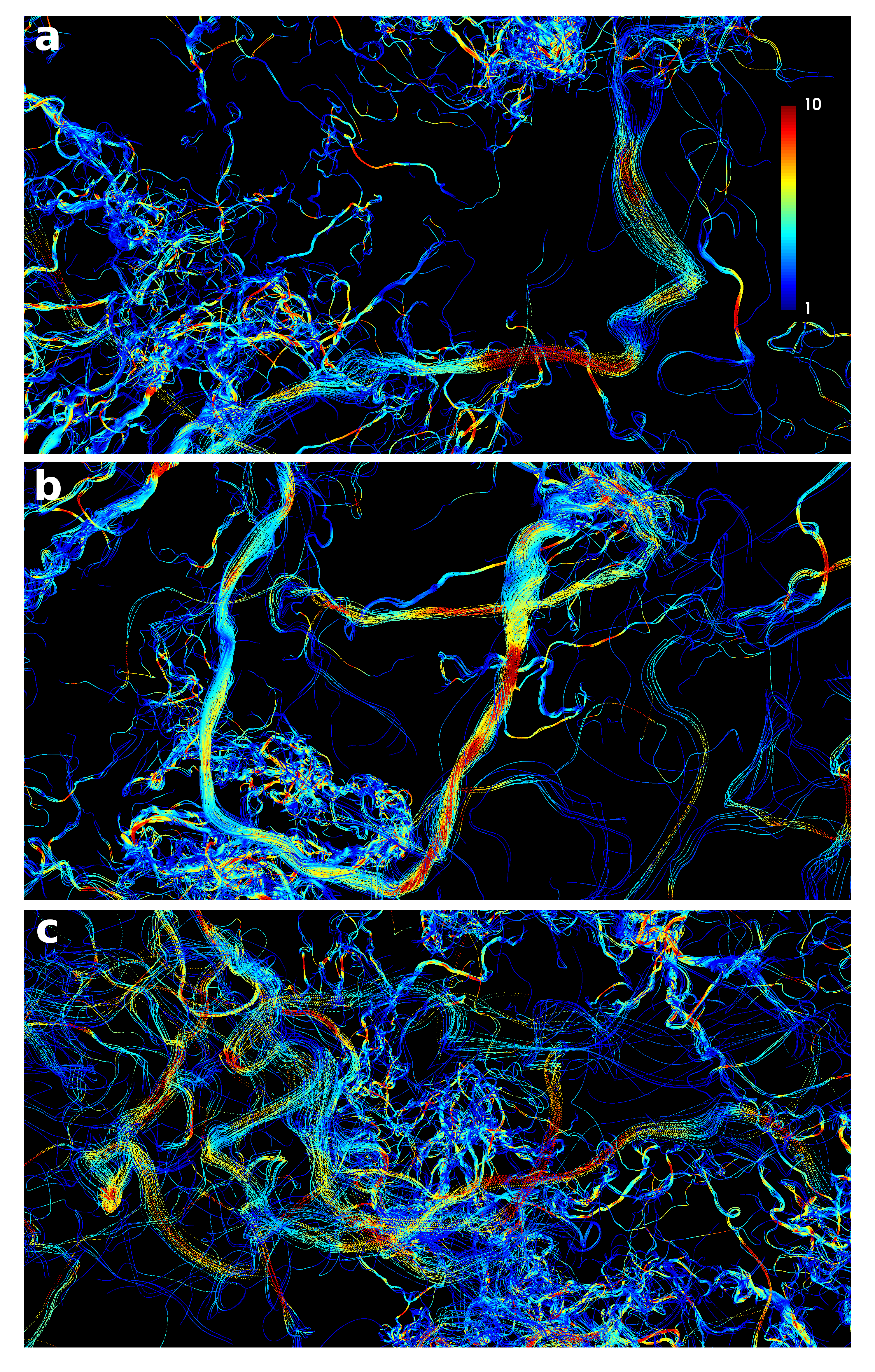}
\caption{A closer look at some VLs, showing a clear preponderance of helical, twisting and torsional bundles, as well as a correlation between higher amplitude vorticity with prolonged coherence lengths in the bundles. Regions of strong vorticity, moreover, appear as localized \textit{vorticity-jets} with low curvature.}
\label{fig:VLSnaps}
\end{figure}

Lastly, in Fig.~\ref{fig:VLSnaps}, we present a few representative snapshots clearly showing the extended coherence of VLs passing through regions of high vorticity magnitude. The VLs clearly bundle in these regions, and are more or less \textit{jet-like}, as has been previously observed~\cite{she1990intermittent,she1991structure,sid2022}. Figs.~\ref{fig:VLSnaps}(b) and \ref{fig:VLSnaps}(c) also show instances of twist and torsion along vortex lines.

\section{Conclusions}
We introduce a method to identify, and characterize the geometry of, vortex lines (VLs)---curves that are everywhere tangent to the vorticity vector---in turbulence. By defining vorticity magnitude thresholds for the start ($\oin$) and end ($\of$) points and tracing forward and backward paths, we obtain the full geometrical structure of individual VLs, together with pertinent physical quantities like velocity, vorticity and curvature along their lengths. By considering ensembles of VLs (of the same start and end vorticity magnitudes as defined by fixing $\oin$ and $\of$) connecting different regions of space, we then study certain statistical features of VLs. 

We show that VLs have exponentially distributed lengths connecting regions with high values of vorticity to regions with mild levels of vorticity. The VL paths are symmetrically distributed in the forward and backward directions. Vorticity decays rapidly along VLs at higher vorticity levels. The turning radii of VLs as measured by the local curvature is shown to be inversely correlated to the vorticity magnitude, which quantifies that high $\omega$ regions are locally jet-like with low curvature---a consequence, we believe, of the vortex stretching mechanism. Finally, we studied the persistence of vorticity along VLs, by considering the average decay of $\omega$ starting from its maximum value along each VL. The large variation in the ensemble reflects the lack of a universal geometry, at least for vorticity decay over length. However, we do observe two regimes corresponding to highly monotonic (from $\omax \to 0.5\omax$) and frequently non-monotonic (from $0.5\omax \to 0.25 \omax$) vorticity decay.

Our work shows that characterizing the complex geometry of VLs in this manner can give new insights regarding their organization and structure, and reinvigorates focus on vortex dynamics. We are further interested in using this method for characterizing important topogical structures like links, writhes and twists, which naturally require curvature and torsion in VLs, and are associated with high helicity density in their neighbourhoods. Moreover, vortex reconnection events, related to high energy dissipation (particularly important in high $Re$ flows), are responsible for VL topology changes. For instance, they lead to the formation of intense vortical structures with high curvatures as well, which are possibly outliers in the statistical observations presented here, characterizing which we leave for future work. This method readily extends to other flows, like shear driven or wall bounded flows, as well as magnetohydrodynamic or superfluid turbulence.





\section{Acknowledgements}
We thank Prof. H.K. Moffatt and Prof. Samriddhi Sankar Ray for insightful discussions during the early stages of this work, and Rajarshi Chattopadhyay for help with obtaining the Johns Hopkins Turbulence Dataset. We acknowledge the ICTS program - Turbulence: Problems at the Interface of Mathematics and Physics (ICTS/TPIMP2020/12). We are grateful for support from the Department of Atomic Energy, Government of India, under Project No. RTI4001.

\section{Declarations}
\subsection*{Approval}
Not applicable.
 
\subsection*{Competing interests}
The authors declare no competing interests.
 
\subsection*{Authors' contributions}
All authors were involved in designing the research. S.K and S.M performed the calculations and S.M made the figures. All authors discussed the results and contributed to writing and editing the manuscript.
 
\subsection*{Funding}
We acknowledge support from the Department of Atomic Energy, Government of India, under Project No. RTI4001.
 
\subsection*{Availability of data and materials}
The data is taken from a publicly available repositiory. Analysis codes can be shared upon request.

\bibliography{sn-bibliography}

\end{document}